\documentclass[10pt,twocolumn,letterpaper,superscriptaddress,prl]{revtex4}

\usepackage{graphicx,amsmath,color,amssymb}
\usepackage[latin1]{inputenc}
\usepackage[sort&compress]{natbib}
\usepackage{epstopdf}
\usepackage{booktabs}
\usepackage[greek,english]{babel}
\bibpunct{}{}{,}{s}{}{,}

\usepackage[dvips]{epsfig}
\usepackage{subfigure,psfrag}
\usepackage{lscape}
\usepackage{varioref}
\usepackage{float}    
\usepackage{tabularx} 
\usepackage[version-1-compatibility]{siunitx}
\definecolor{darkblue}{rgb}{0,0,.4}
\definecolor{darkred}{rgb}{0.6,0,0}

\renewcommand\eqref[1]{Eq.~(\ref{#1})}

\newcommand{\rr}{\mathbf{r}}
\newcommand{\RR}{\mathbf{R}}
\newcommand{\me}[3]{\left< #1 \left| #2 \right| #3 \right>}
\newcommand{\ovI}[2]{\left< #1 | #2 \right>}

\usepackage[bookmarksopen=true, 
citecolor=blue,
colorlinks=true, 
linkcolor=blue]{hyperref} 

\begin{document}

\title{Unraveling the mesoscopic character of quantum dots in nanophotonics}

\author{P.~Tighineanu}
\email{petru.tighineanu@nbi.ku.dk}
\affiliation{Niels Bohr Institute,\ University of Copenhagen,\ Blegdamsvej 17,\ DK-2100 Copenhagen,\ Denmark}

\author{A.~S.~S\o rensen}
\affiliation{Niels Bohr Institute,\ University of Copenhagen,\ Blegdamsvej 17,\ DK-2100 Copenhagen,\ Denmark}

\author{S.~Stobbe}
\affiliation{Niels Bohr Institute,\ University of Copenhagen,\ Blegdamsvej 17,\ DK-2100 Copenhagen,\ Denmark}

\author{P.~Lodahl}
\email{lodahl@nbi.ku.dk}
\homepage{http://www.quantum-photonics.dk/}
\affiliation{Niels Bohr Institute,\ University of Copenhagen,\ Blegdamsvej 17,\ DK-2100 Copenhagen,\ Denmark}

\date{\today}

\small

\begin{abstract}
We provide a microscopic theory for semiconductor quantum dots that explains the pronounced deviations from the prevalent point-dipole description that were recently observed in spectroscopic experiments on quantum dots in photonic nanostructures. At the microscopic level the deviations originate from structural inhomogeneities generating a large circular quantum current density that flows inside the quantum dot over mesoscopic length scales. The model is supported by the experimental data, where a strong variation of the multipolar moments across the emission spectrum of quantum dots is observed. Our work enriches the physical understanding of quantum dots and is of significance for the fields of nanophotonics, quantum photonics, and quantum-information science, where quantum dots are actively employed.
\end{abstract}


\maketitle

\begin{figure}[t!]
	\includegraphics[width=\columnwidth]{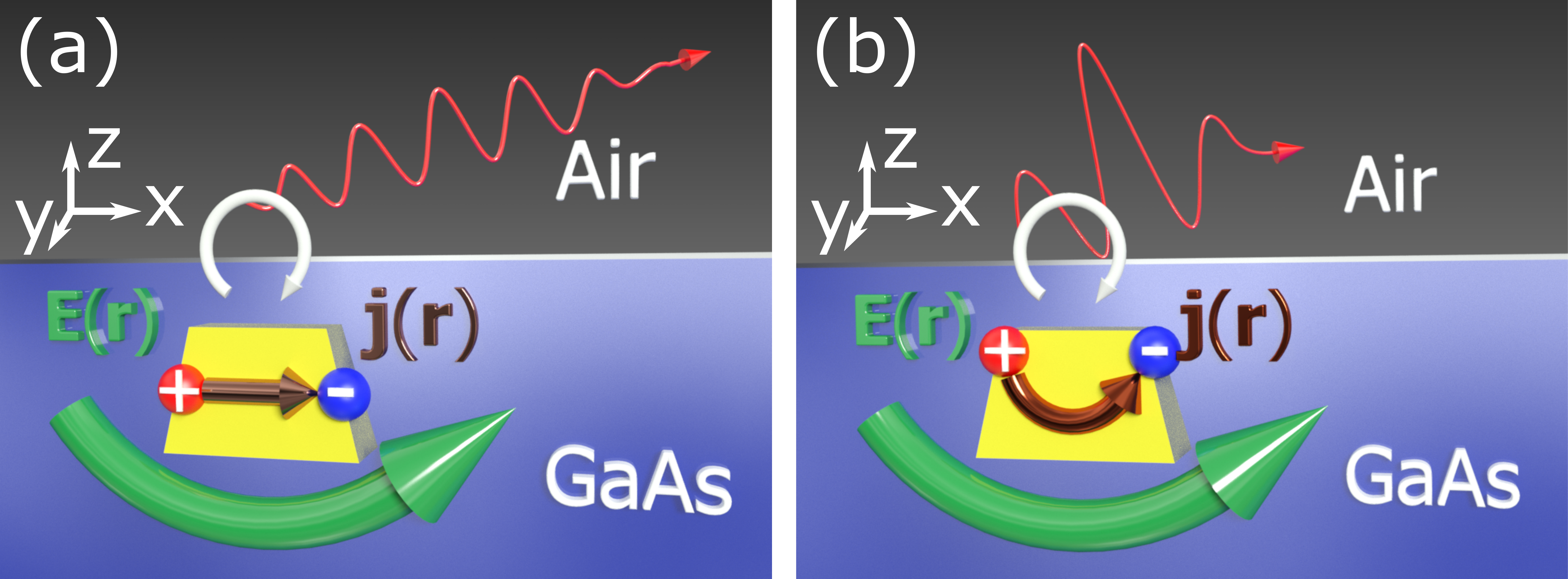}
	\caption{ \label{fig:QuantumBanana} Unraveling the mesoscopic character of QDs in the vicinity of a GaAs-air interface. The presence of the interface breaks the parity symmetry of the environment in the $z$-direction. Since reflections occur at the interface (the circular white arrow), the imaginary part of the electric field $\mathbf{E}(\rr)$ generated by the electric-dipole component, which triggers spontaneous emission, inherits this lack of symmetry and is curved (indicated by the green arrow). (a) In the dipole approximation, the QD microscopic current $\mathbf{j}(\rr)$ (brown arrow) perceives only the parallel component but not the out-of-plane component (the "curvature") of the electromagnetic field at its position. (b) In In(Ga)As self-assembled QDs, the current density flows along a curved path that resembles the shape of the field environment thereby exchanging energy more efficiently with it. As a consequence, the spontaneous-emission decay rate is enhanced and the photons (red arrows) are emitted at a faster rate compared to the case in (a).}
\end{figure}

Semiconductor quantum dots (QDs) are compatible with semiconductor technology and exhibit optically active transitions with high quantum efficiency, which renders them promising single-photon sources for a range of solid-state quantum-optical devices~\cite{lodahl13}. For instance, strong coupling between a QD and a cavity photon~\cite{yoshie04} or near-unity coupling to a photonic-crystal waveguide have been demonstrated~\cite{arcari14}. Furthermore, QDs have been proposed as the enabling part of highly efficient solar cells\cite{atwater10} and as central nodes in future quantum-information systems with entangled stationary and flying quantum bits~\cite{gao12}. All of these applications require a profound understanding of their optical properties. The latter was recently challenged by the observation that QDs can break the dipole approximation and was explained by a phenomenologically defined mesoscopic moment~\cite{andersen10}, which probes the multipolar electromagnetic vacuum field~\cite{tighineanu14}. So far, the physical origin of these mesoscopic quantum effects has not been established. Here we present a microscopic theory of the QD wavefunctions that provides physical insight into the mesoscopic character of QDs. We find that the microscopic quantum current density flows along a curved path inside the QD, cf. Fig.~\ref{fig:QuantumBanana}(b), which generates large electric and magnetic moments leading to light-matter interaction of both electric and magnetic character~\cite{sersic11}. The mesoscopic moment, which includes the magnetic light-matter interaction, can be tuned over orders of magnitude by controlling the size and shape of QDs. The theoretical model is supported by comparing to experimental data on time-resolved spectroscopy of QDs positioned near a dielectric interface in a Drexhage geometry~\cite{novotny12} (cf. Fig.~\ref{fig:QuantumBanana}), where a strong variation of the mesoscopic moment with emission energy is found for the case of In(Ga)As self-assembled QDs.

A central quantity describing the optical transition from the excited state $\Psi_e$ to the ground state $\Psi_g$ of a QD is the dipole moment $\boldsymbol \mu = (e/m_0)\left<\Psi_g | \hat{\mathbf{p}} | \Psi_e \right>$, where $e$ and $m_0$ are the elementary charge and electron mass, respectively. We consider the $x$-polarized dipole moment of the exciton $\boldsymbol \mu = \mu \hat{\mathbf{x}}$, as sketched in Fig.~\ref{fig:QuantumBanana}, where $\hat{\mathbf{x}}$ is the Cartesian unit vector. Until recently, $\mu$ was the only QD property used to describe the interaction with light. Recent experimental studies of spontaneous emission from QDs near a silver interface are, however, incompatible with such a description~\cite{andersen10} and a more extended description is required. So far, the decay dynamics beyond the dipole approximation was described by a phenomenological mesoscopic moment $\Lambda=(e/m_0)\left<\Psi_g | x\hat{p}_z |\Psi_e \right>$~\cite{tighineanu14}. Combined with the microscopically well-understood dipole moment, this quantity accounts for the interaction with light caused by the extended mesoscopic nature of QDs. Previous theories have investigated mesoscopic effects at the level of the QD spatial extent and symmetry, and have discarded their atomistic nature because the unit cells are small compared to the wavelength of light~\cite{goupalov03,sugawara95,thranhardt02,ahn03,zurita01,zurita02,stobbe12,kristensen13}. These approaches fail to explain the large mesoscopic moment $\Lambda$ observed experimentally. In the present work we show that, surprisingly, the atomistic character plays a crucial role and explains the physical origin of the mesoscopic moment. We find that the latter is caused by the change in the periodicity of the underlying crystal lattice in the QD. Since Bloch functions with different periodicities cannot remain in phase throughout the QD, this necessarily leads to a phase gradient and a resulting current in the growth direction of the QD, which gives rise to the mesoscopic moment, cf. Fig.~\ref{fig:QuantumBanana}(b). As a result, QDs sense electromagnetic-field gradients and probe electric- and magnetic-field fluctuations simultaneously. To obtain an optimized light-matter coupling, the electromagnetic environment must be shaped similarly to the flow of the QD current, as sketched in Fig.~\ref{fig:QuantumBanana}(b), where we exemplify the case of a GaAs-air interface. The experimental data demonstrating this effect are presented in the following.

\begin{figure}[t!]
	\includegraphics[width=\columnwidth]{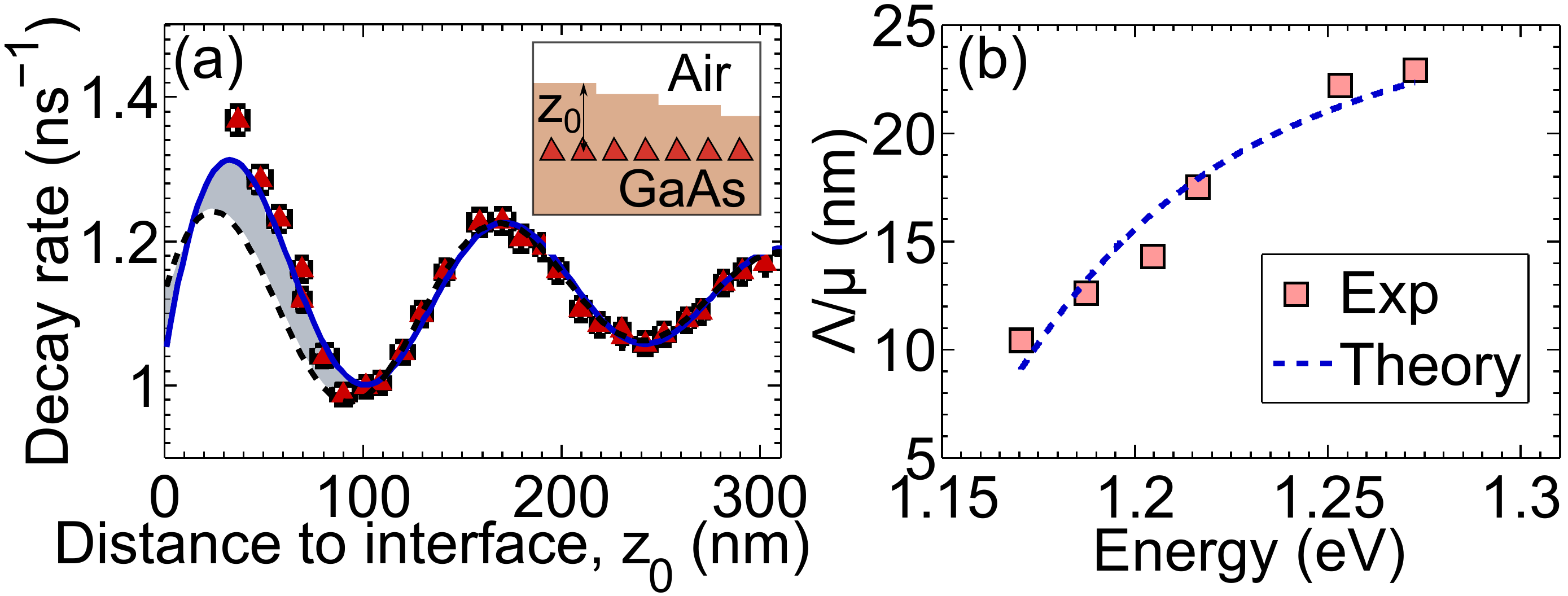}
	\caption{ \label{fig:InterfaceData} Observation of deviations from dipole theory for QDs near an interface. (a) Measured decay rates versus distance $z_0$ to the GaAs-air interface (data points) at an energy of \SI{1.27}{\electronvolt}. The dipole (multipolar) theory is indicated by the black dashed (blue solid) line. A refractive index $n=3.5$ of GaAs was used. The inset is a schematic illustrating the sample geometry.  (b) Extracted mesoscopic strength $\Lambda/\mu$ over the emission spectrum of QDs (red squares) along with the prediction of the theoretical model (blue dashed line) assuming that the QDs have a fixed in-plane size and only the height varies, cf. Supplementary Information.}
\end{figure}

Figure~\ref{fig:InterfaceData}(a) displays the measured spontaneous-emission decay rates for ensembles of QDs that are placed at different distances to a GaAs-air interface. The experimental data were published in Refs.~\citenum{johansen08,stobbe09}, where the data recorded at distances above $\SI{75}{\nano\meter}$ were employed to reliably extract the dipole contribution to the spontaneous-emission process. We observe a systematic deviation from dipole theory at distances below $\sim \SI{75}{\nano\meter}$, which was previously speculated to be a result of enhanced loss processes at the etched interfaces~\cite{johansen08}. These deviations can instead be explained by the contributions from the mesoscopic moment $\Lambda$ to the light-matter interaction strength, and we are able to quantitatively reproduce the functional dependence observed in the experiment for all emission energies, as exemplified in Fig.~\ref{fig:InterfaceData}(a), see Supplementary Information for details. The extracted mesoscopic strength $\Lambda/\mu$ increases with emission energy and varies from 10 to \SI{23}{\nano\meter} over the inhomogeneously broadened emission spectrum, cf. Fig.~\ref{fig:InterfaceData}(b), which reflects a pronounced dependence on QD size.
In the following we develop a theory that quantitatively explains the experimental data of Fig.~\ref{fig:InterfaceData}(b).

Numerous band structure models have been proposed for QDs ranging from continuum approaches (e.g., the multiband k.p theory~\cite{luttinger55}), which discard the QD atomistic nature and consider only the macroscopic potential, to atomistic models (e.g., the empirical pseudopotential theory~\cite{bester09}), which simulate the contribution of every single atom comprising the QD. In the present work, it suffices to consider the simplest band structure model, i.e., the two-band effective-mass theory. We show that, by including a small extension of it, the mesoscopic nature of QDs can be explained remarkably well. In this picture, the QD single-particle wavefunction $\Psi(\rr)$ can be written as a product of a periodic Bloch function $u(\rr)$, which captures the properties on the length scale of the crystal unit cell, and a slowly varying envelope $\psi(\rr)$ inheriting the size and symmetry of the QD potential, i.e., $\Psi_i(\rr) = u_i(\rr)\psi_i(\rr)$ with $i=\{g,e\}$ corresponding to either the ground or excited state, respectively. We consider the $x$-polarized exciton with the valence-band heavy-hole Bloch function $u_g=u_x$~\cite{harrison05} inheriting the atomic $p_x$ symmetry. The standard textbook approach for evaluating the transition dipole moment $\mu$ is to assume that the envelope function varies slowly over a unit cell so that $\mu$ can be written as a product of the Bloch matrix element $p_{cv}$ and a three-dimensional overlap integral between the envelope functions, i.e., $\mu=(e/m_0)\left< u_x \psi_g | \hat{p}_x | u_e \psi_e \right> \approx (e/m_0) p_{cv} \left< \psi_g | \psi_e \right>$, where $p_{cv}=V_\mathrm{UC}^{-1} \int_\mathrm{UC}d^3\rr u_x^* \hat{p}_x u_e$ is given by an integral over the unit cell with $V_\mathrm{UC}$ being the unit-cell volume. In other words, the transition dipole moment is primarily a unit-cell effect and is marginally affected by the envelope functions, as their overlap is normally close to unity~\cite{johansen08}. Importantly, the large mesoscopic strength $\Lambda/\mu$ observed experimentally (see Fig.~\ref{fig:InterfaceData}(b)) cannot be reproduced by a similar calculation, which leads to
\begin{equation}
\begin{split}
\Lambda &= \frac{e}{m_0}\left[ \me{\psi_g}{x}{\psi_e}\me{u_x}{\hat{p}_z}{u_e}_\mathrm{UC} + \ovI{\psi_g}{\psi_e}\me{u_x}{x\hat{p}_z}{u_e}_\mathrm{UC} \right. \\
&+ \left. \me{\psi_g}{x\hat{p}_z}{\psi_e}\ovI{u_x}{u_e}_\mathrm{UC} + \me{\psi_g}{\hat{p}_z}{\psi_e}\me{u_x}{x}{u_e}_\mathrm{UC} \right],
\end{split}
\end{equation}
where $<>_\mathrm{UC}\equiv V_\mathrm{UC}^{-1}\int_\mathrm{UC} \mathrm{d}^3\rr$ denotes integration over a unit cell. The first three contributions vanish for symmetry reasons. The fourth contribution is vanishingly small and  does not scale with the QD size: for Gaussian envelopes allowing for realistic mutual displacements of 1--\SI{2}{\nano\meter} between the electron and the hole in the growth direction (note that the integral vanishes in the absence of such a displacement) we estimate $\Lambda/\mu \sim 10^{-4}\si{\nano\meter}$. This suggests that the large mesoscopic strength $\Lambda/\mu \sim10$--\SI{20}{\nano\meter} observed experimentally cannot be explained  solely by the  envelope wavefunctions. In the following we show that structural gradients at the nanoscopic crystal-lattice length scale can explain the effect.

\begin{figure}[t!]
	\includegraphics[width=0.45\textwidth]{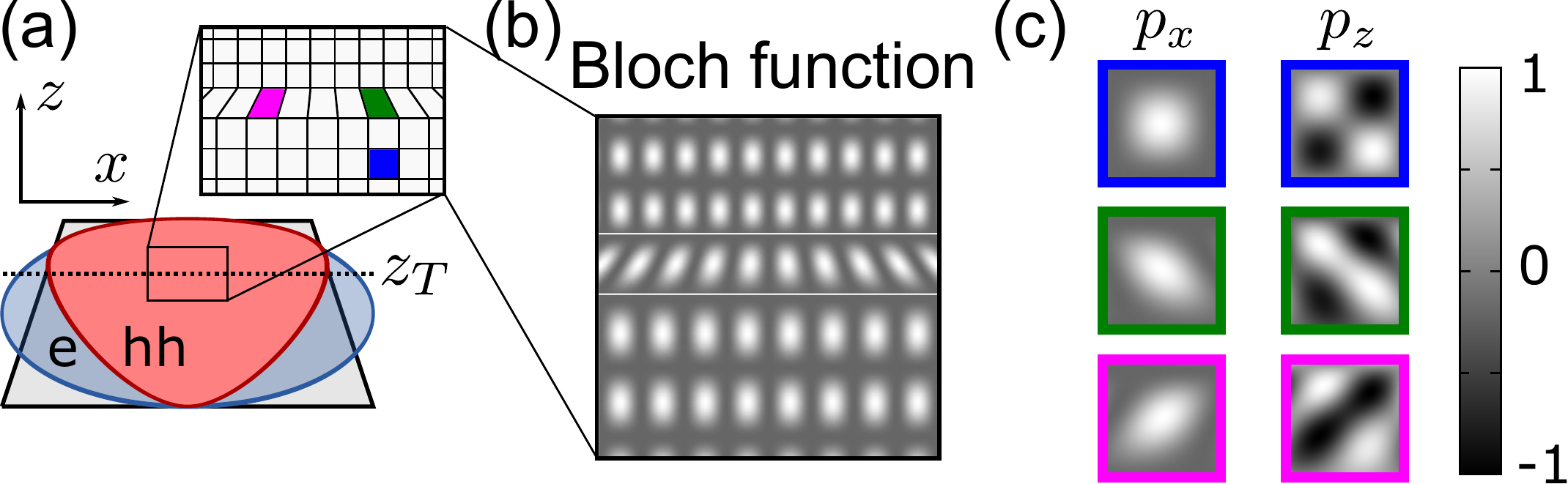}
	\caption{ \label{fig:LatticeChange} Sketch illustrating the microscopic model for mesoscopic QDs. (a) The atomic lattice inside the QD is assumed to change periodicity at the position $z=z_T$. (b) Sketch of how the Bloch function $u_x^2$ of the atomic lattice varies spatially inside the QD.  (c) Illustration of the matrix elements $\left< p_x \right> \equiv \left<u_x | \hat{p}_x | u_e \right>$ and $\left< p_z \right> \equiv \left<u_x | \hat{p}_z | u_e \right>$ for the three colored unit cells in (a). The symmetry of the integrand is broken in the transition region around $z=z_T$ giving rise to pronounced mesoscopic effects.}
\end{figure}

It is often assumed that solid-state emitters have a homogeneous chemical composition, which renders substantial simplifications in the computation of the wavefunctions. In particular, the homogeneity justifies the use of bulk-material Bloch functions, and only the slowly varying envelopes describe the properties of the nanostructure. This assumption works excellently for quantum wells and lattice-matched QDs, where the structures are either strain free or pseudomorphically grown on the substrate material. As a result, the wavefunctions are confined to a chemically homogeneous region of space. InAs QDs are grown by self-assembly induced by strain relaxation, which unavoidably leads to the generation of chemical gradients at the crystal-lattice level. In particular, large lattice-constant shifts were observed in the growth direction of QDs~\cite{eisele08,bruls02}. This limits the applicability of the standard envelope-function theories and, in particular, of the effective-mass formalism. A complete theory encompassing the spatial position and symmetry of every single atom comprising the QD would generally be required. Remarkably, the essential physics of the mesoscopic light-matter interaction can be captured by only a minor extension of the effective-mass theory. We assume that the lattice periodicity changes at a certain position $z=z_T$ along the QD height by an amount $\Delta a_l=\SI{110}{\pico\meter}$ at a central value $a_l=\SI{605}{\pico\meter}$ as found experimentally in Ref.~\citenum{eisele08}, see Fig.~\ref{fig:LatticeChange}(a). This corresponds to a relative lattice-constant shift of 18\%, which is strain induced and is substantially larger than the lattice-constant mismatch between InAs and GaAs of 7\%. We note that, in general, the lattice periodicity changes twice: first it is expanded at the QD base (GaAs-In(Ga)As transition) before being shrunk back at the QD tip (In(Ga)As-GaAs transition). Since the exciton is spatially confined near the tip where the indium concentration is highest~\cite{bruls02}, we only consider the second transition region. The Bloch functions change periodicity as well, cf. Fig.~\ref{fig:LatticeChange}(b), and we model this by expanding them in a Fourier series with a position-dependent lattice wavevector $k_l(z)$
\begin{equation}
\begin{split}
u_x(\rr) &= \sum_m a_m(y,z) \sin[mk_l(z)x]\\
u_e(\rr) &= \sum_n b_n(y,z) \cos[nk_l(z)x].
\end{split}
\label{eq:blochAnzatz}
\end{equation}
This Ansatz ensures opposite parity of the conduction- and valence-band Bloch functions along $x$. Furthermore, we implicitly assume the shape of the Bloch functions to remain the same, and only their periodicity to vary spatially. Now we return to the evaluation of the mesoscopic moment and separate the slowly- and rapidly-varying contributions as
\begin{equation}
\Lambda = \frac{e}{m_0}\sum_{q=1}^{N}\psi_{g}^*(\RR_q)X_q\psi_e(\RR_q)\int_\mathrm{UC}\mathrm{d}^3\rr u_x^*(\rr) \hat{p}_z u_e(\rr),
\label{eq:Lambda_in}
\end{equation}
where $\RR_q$ denotes the position of the $q$-th unit cell and $N$ is the total number of unit cells in the QD. In a homogeneous region of the QD (the blue unit cell in Fig.~\ref{fig:LatticeChange}(a)) the unit-cell integrand of \eqref{eq:Lambda_in} is odd in $x$- and $z$-directions, cf. Fig.~\ref{fig:LatticeChange}(c), which leads to a vanishing integral. However, in the transition region around $z=z_T$ strong gradients are present, which destroy the parity of the integrand (see the pink and green unit cells in Fig.~\ref{fig:LatticeChange}(a,c)) and generate a substantial contribution to $\Lambda$.

\begin{figure}[t!]
	\includegraphics[width=0.48\textwidth]{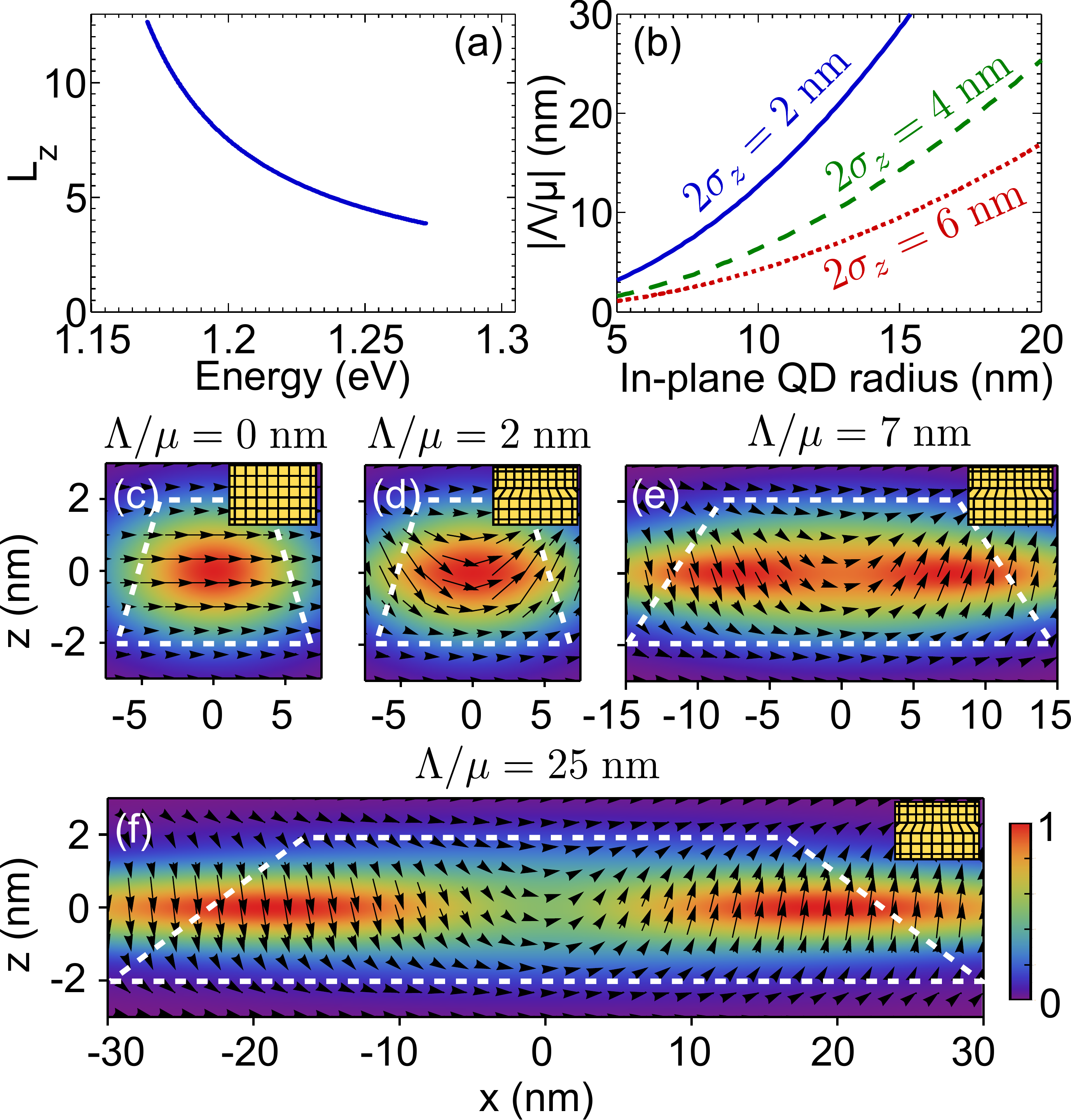}
	\caption{ \label{fig:Current} The mesoscopic strength and the associated current density running through the QD. (a) Spectral dependence of the QD height predicted by the theoretical model. (b) Mesoscopic strength as a function of the in-plane size of the QD for three fixed QD heights. (c)-(f) Plot of quantum current densities for various QD geometries. (c) Homogeneous crystal lattice where the current flow is uniform and points in the direction of the dipole moment. (d) Inhomogeneous lattice for a QD radius of \SI{5}{\nano\meter} giving rise to a non-uniform current flow following a curved path. The QD height is \SI{4}{\nano\meter}. (e),(f) Same as (d) but for QD radii of 10 and \SI{20}{\nano\meter}, respectively. In (c)-(f), both the length of the arrows and the color scale indicate the magnitude of the flow and the direction of the arrows indicate  the pointwise direction of the flow. The dashed white line sketches the position and orientation of the QD.}
\end{figure}

In the following, we calculate the mesoscopic moment $\Lambda$ and show that its magnitude is sensitive to the QD geometry. With the Ansatz in \eqref{eq:blochAnzatz} we first compute the dipole Bloch matrix element $\left<\hat{p}_x\right>$ and then evaluate $\Lambda$, which can be expressed in terms of $\left<\hat{p}_x\right>$ because the $x$- and $z$-derivatives of the Bloch functions yield similar results, cf. Eqs.~(\ref{eq:blochAnzatz}--\ref{eq:Lambda_in}). The resulting expression for $\Lambda/\mu$ reads
\begin{equation}
\frac{\Lambda}{\mu} = \frac{1}{k_l}\frac{\me{\psi_g(\rr)}{x^2\left[\partial_z k_l(z)\right]}{\psi_e(\rr)}}{\ovI{\psi_g(\rr)}{\psi_e(\rr)}}.
\label{eq:Lambda_mu_fin}
\end{equation}
We have thus been able to express a crystal-lattice effect in terms of the slowly varying envelope functions. The mesoscopic strength scales quadratically with the in-plane size of the QD, $\Lambda/\mu \sim L_r^2$, because the term $\me{\psi_{g}}{x^2\left[\partial_z k_l(z)\right]}{\psi_e}$ contains the variance of the exciton wavefunction in the $x$-direction. Moreover, it increases with decreasing QD height, $\Lambda/\mu \sim L_z^{-1}$, since in shallow QDs the relative importance of the lattice-constant transition region is increased. We use \eqref{eq:Lambda_mu_fin} to model the spectral dependence of $\Lambda/\mu$, see Fig.~\ref{fig:InterfaceData}(b), where only the height of QDs is assumed to vary across the spectrum while the in-plane size remains constant. This assumption is supported by studies of size and shape performed on self-assembled QDs~\cite{moison94}, where a small relative distribution of the in-plane QD size is observed. By mapping the quantization energy to the QD size, we are able to extract a QD height that varies from \SI{11}{\nano\meter} to \SI{3}{\nano\meter} across the inhomogeneously broadened spectrum, cf. Fig.~\ref{fig:Current}(a), which agrees well with the values obtained from atomic-force microscopy measurements~\cite{stobbe09}. In the present work we assume $z_T=0$, i.e., the transition region is situated at the center of the QD slowly varying envelopes. The details of the calculation are given in the Supplementary Information. From Fig.~\ref{fig:InterfaceData}(b) we conclude that QDs with larger emission energy have larger mesoscopic strengths because they are shallow so that a large part of the excitonic wavefunction is affected by the lattice inhomogeneity.

The existence of deviations from the dipole theory even in simple weakly confining dielectric structures (cf. Fig.~\ref{fig:InterfaceData}) points to their relevance in any nanophotonic structure that breaks parity symmetry along the QD height. Motivated by these findings, we derive a simplified analytic relation between the QD geometry and the mesoscopic strength for in-plane rotationally symmetric Gaussian slowly varying envelopes
\begin{equation}
\frac{\Lambda}{\mu} = -\frac{\Delta a_l}{a_l}\sqrt{\frac{1+\xi_z}{4\pi}} \frac{\sigma_r^2}{\sigma_z},
\label{eq:Lambda_Mu_Gauss}
\end{equation}
where $\sigma_z$ is the height (HWHM) of the electron envelope, $\sigma_r$ the QD radius, $\Delta a_l/a_l$ the relative lattice-constant shift and $\xi_z\approx 5$ is the ratio between the electron and hole effective masses~\cite{cardona10}. We plot the mesoscopic strength as a function of the in-plane radius for three fixed heights in Fig.~\ref{fig:Current}(b). The largest mesoscopic strengths are achieved by shallow and wide (disk-shaped) QDs. For instance, taking a relative extreme case of a height of $\sigma_z = \SI{2}{\nano\meter}$ and a radius of $\sigma_r=\SI{30}{\nano\meter}$ yields a mesoscopic strength as large as $\Lambda/\mu = \SI{150}{\nano\meter}$, which is an order of magnitude larger than the values observed in experiments so far. Such QDs would constitute a mesoscopic entity in which mesoscopic effects may dominate the light-matter interaction strength. For instance, a QD with $\Lambda/\mu = \SI{150}{\nano\meter}$ placed in front of a silver mirror would exhibit a Purcell factor that is nearly 100 times larger than the case of a point-dipole source. Aside from this, such QDs may be extremely efficient at interfacing both electric and magnetic degrees of freedom also in structures that conserve parity symmetry, such as photonic-crystal cavities and waveguides, owing to the substantial increase of second-order light-matter-interaction processes that are weak for current In(Ga)As QDs.

Knowledge about the quantum-mechanical wavefunctions allows computing the current density $\mathbf{j}_\mathrm{QD}(\rr)$ flowing through the QD. We define the latter by comparing the interaction Hamiltonian $\hat{H}_\mathrm{int}=(e/m_0)\mathbf{A}\cdot\hat{\mathbf{p}}$, where $\mathbf{A}$ is the vector potential, to the classical particle-field interaction Hamiltonian $H_\mathrm{int}= \mathbf{A}(\rr) \cdot \mathbf{j}(\rr)$~\cite{novotny12}. The quantum-mechanical current density can therefore be written as $\mathbf{j}_\mathrm{QD}(\rr) = (e/m_0) \Psi_g(\rr) \hat{\mathbf{p}} \Psi_e(\rr)$ or
\begin{equation}
\mathbf{j}_\mathrm{QD}(\rr) = \frac{e}{m_0}\left[ \Psi_g\hat{p}_x\Psi_e \hat{\mathbf{x}} + \Psi_g\hat{p}_z\Psi_e \hat{\mathbf{z}} \right].
\label{eq:j_def}
\end{equation}
In the following, Gaussian slowly varying envelopes are used to model the current density. In QDs with a homogeneous crystal lattice and thus negligible mesoscopic moment, the current density flows only along the direction of the dipole moment because there are no gradients in the $z$-direction and the second term from the right-hand side of \eqref{eq:j_def} vanishes (see Fig.~\ref{fig:Current}(c)). Note that for simplicity we ignore the modulation of $\mathbf{j}_\mathrm{QD}$ by the periodic Bloch functions in Fig.~\ref{fig:Current}. The presence of lattice inhomogeneities changes the flow dramatically because strong gradients in the $z$-direction arise. The current density flows along a curved path as illustrated in Figs.~\ref{fig:Current}(d-f), conferring pronounced mesoscopic properties to QDs. The wider the QD is, the sharper the transverse oscillations of the current are and the larger $\Lambda/\mu$ is. This effect offers the possibility to enhance (diminish) the light-matter interaction by placing QDs in environments where the electric vacuum field exhibits gradients with the same (opposite) sign, see also Fig.~\ref{fig:QuantumBanana}. This opens new opportunities for designing efficient light-matter interfaces that exploit mesoscopic effects to enhance the interaction. Aside from the local light-matter coupling strength, other degrees of freedom could be potentially tailored by exploiting the mesoscopic interaction, such as the photon-emission directionality or polarization.

In conclusion, we have developed a novel microscopic model that successfully explains the large mesoscopic strengths of In(Ga)As QDs observed experimentally. We find the effect to be governed by the lack of symmetry of the nanoscopic crystal lattice and scaled by the extended mesoscopic size of the QD. The microscopic current density oscillates along a non-trivial curved path and can be expressed as a superposition between electric-dipole, magnetic-dipole, and electric-quadrupole moments. This mesoscopic current is generated at the unit-cell level in analogy to the generation of currents in macroscopic systems. Our work deepens the physical understanding of semiconductor QDs and we therefore expect it to be of significance for the active fields of solid-state quantum electrodynamics and quantum-information processing, where efficient quantum interfaces between QDs and light are exploited.

\section*{Acknowledgements}
We thank P.~T.~Kristensen and M.~L.~Andersen for valuable discussions. We gratefully acknowledge the financial support from the Danish Council for Independent Research (natural sciences and technology and production sciences), the European Research Council (ERC consolidator grants "ALLQUANTUM" and "QIOS"), and the Carlsberg Foundation.

\bibliographystyle{nature}
\bibliography{bibliography}

\clearpage

\section*{Supplementary Information}
\begin{figure}[H]
	\centering
	\includegraphics[width=0.65\columnwidth]{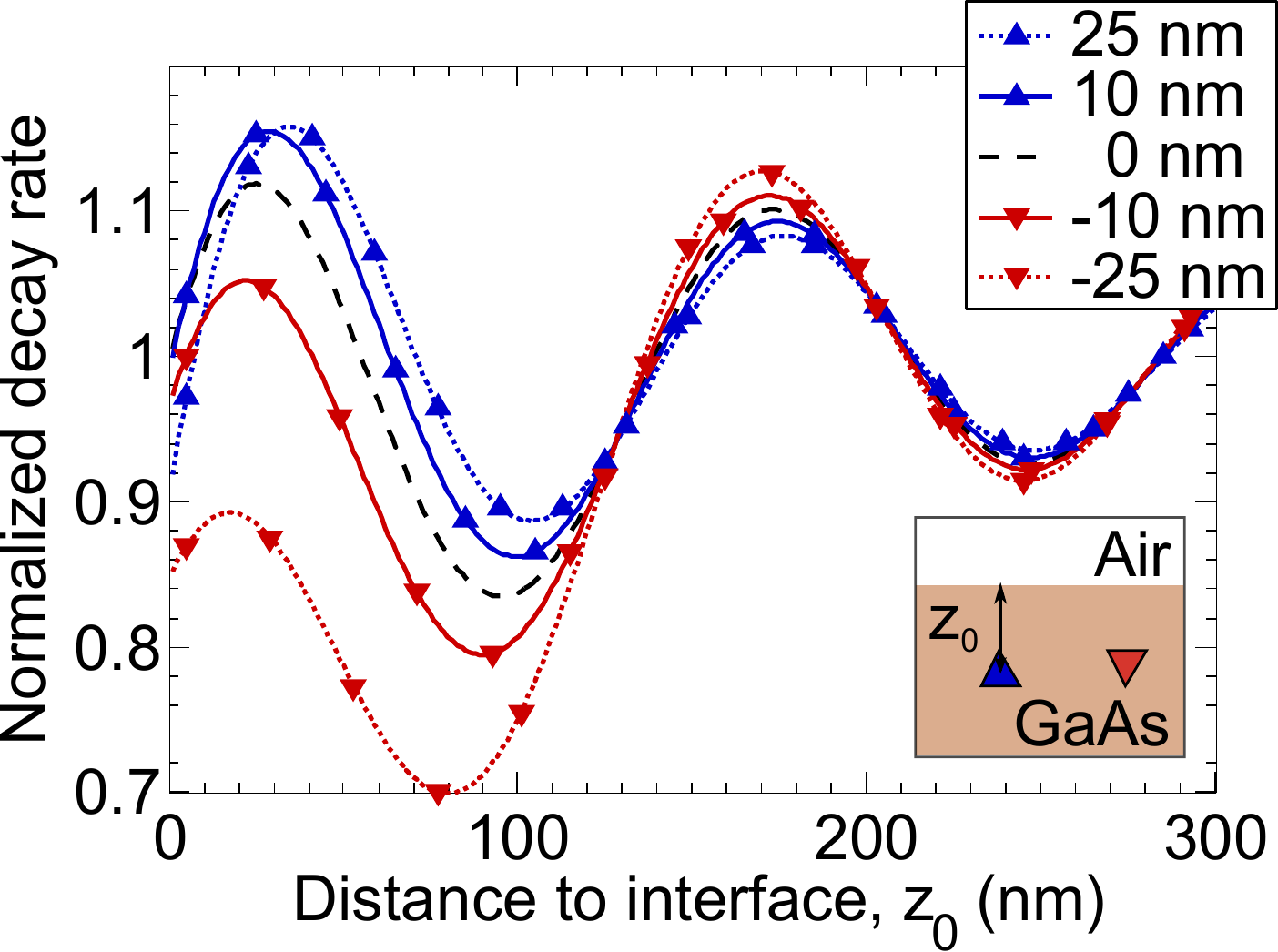}
	\caption{ \label{fig:InterfaceTheory} Calculated spontaneous-emission decay rates as a function of distance to a GaAs-air interface for various mesoscopic strengths. The vanishing mesoscopic strength corresponds to the dipole theory. The decay rates are normalized to the decay rate in homogeneous GaAs. A refractive index $n=3.5$ of GaAs and an emission energy of $E=\SI{1.24}{\electronvolt}$ (wavelength $\lambda=\SI{1}{\micro\meter}$) were employed.}
\end{figure}

Here we outline the details of the theory used to analyze the experimental data of Fig.~\ref{fig:InterfaceData}. We start with Fig.~\ref{fig:InterfaceData}(a), where the spontaneous-emission decay rates as a function of the distance $z_0$ to the air interface are plotted. The decay rate beyond the dipole approximation can be decomposed into three decay channels $\Gamma = \Gamma^{(0)} + \Gamma^{(1)} + \Gamma^{(2)}$ with~\cite{tighineanu14}
\begin{equation}
\begin{split}
\Gamma^{(0)}(\rr_0) &= C\mu^2 \Im\left\{ G_{xx}(\rr_0,\rr_0) \right\}\\
\Gamma^{(1)}(\rr_0) &= 2C\Lambda\mu \left. \partial_x \Im\left\{ G_{zx}(\rr,0) \right\}\right|_{z=z_0}\\
\Gamma^{(2)}(\rr_0) &= C\Lambda^2 \left. \partial_x \partial_x' \Im\left\{ G_{zz}(\rr,\rr') \right\} \right|_{z=z'=z_0},
\end{split}
\label{eq:Gamma_orders}
\end{equation}
where $G_{ij}(\rr,\rr')$ is the electromagnetic Green's tensor, which is evaluated with the help of Ref.~\citenum{paulus00}, $C=2e^2/\epsilon_0\hbar m_0^2c_0^2$, $e$ is the elementary charge, $m_0$ the electron mass, $\epsilon_0$ the vacuum permittivity, and $c_0$ the vacuum speed of light.  We set $\Lambda$ and $\mu$ as free parameters and fit the experimental data with the resulting dependence plotted in Fig.~\ref{fig:InterfaceData}(a). It should be mentioned that a data point observed at a distance of \SI{20}{\nano\meter} from the GaAs-air interface is not shown in Fig.~\ref{fig:InterfaceData}(a) and is omitted from the analysis because it shows a much higher decay rate and lower photoluminescence intensity, which is likely to be caused by nonradiative tunneling of the QD charge carriers to surface states~\cite{stobbe09}. A phenomenological distance-independent loss rate is added to $\Gamma$ to account for intrinsic nonradiative decay channels within the QD but is found to be negligibly small in the present analysis. This procedure is used independently for every emission energy resulting in the data points in Fig.~\ref{fig:InterfaceData}(b). The dipole-theory fit is performed by setting $\Lambda=0$ and excluding the first six data points from the analysis. For completeness, we plot distance-dependent decay rates for various mesoscopic strengths in Fig.~\ref{fig:InterfaceTheory}. It can be seen that both the amplitude and the phase of the oscillations are substantially affected at distances smaller than $\sim \SI{100}{\nano\meter}$ from the interface, which allows extracting the mesoscopic strength from the experimental data. It is interesting to note that the effect is more dramatic for QDs flipped upside down for which $\Lambda/\mu < 0$ as indicated by the red curves in Fig.~\ref{fig:InterfaceTheory}. In Ref.~\citenum{andersen10} this effect was used to demonstrate the existence of the mesoscopic moment.

In the following we present the analysis of the experimental data from Fig.~\ref{fig:InterfaceData}(b) using the theoretical model developed in the paper. In the experiment we map the mesoscopic strength as a function of emission energy $E$. Equation~(\ref{eq:Lambda_mu_fin}) implicitly describes the mesoscopic strength versus the QD size, as explained in the text. We therefore convert the QD size into a quantization energy and in this regard make several assumptions. First, we consider the inhomogeneously broadened spectrum to be caused only by the random distribution of the size (and consequently of the quantization/emission energy) of QDs. Other parameters such as strain distribution or chemical composition are considered constant over the emission spectrum. Second, we assume that only the height of the QDs contributes to the quantization energy, since atomic-force microscopy measurements show that the height of self-assembled In(Ga)As is generally much smaller than the in-plane size~\cite{bruls02}. Third, we assume a sharp transition in the lattice constant $\partial_z k_l = \Delta k_l \delta(z-z_T)$ for simplicity. This approximation is excellent because $\partial_z k_l$ is multiplied with a slowly varying integrand in \eqref{eq:Lambda_mu_fin}, see also Fig.~\ref{fig:LatticeChange}(a). 
Fourth, we assume disk-shaped wavefunctions that can be decomposed into in-plane $\Phi$ and out-of-plane $\phi$ components, i.e., $\psi(\rr) = \Phi(x,y)\phi(z)$. With this, \eqref{eq:Lambda_mu_fin} can be rewritten
\begin{equation}
\frac{\Lambda}{\mu} = - \frac{\Delta a_l}{a_l} \frac{\me{\Phi_g}{x^2}{\Phi_e}}{\ovI{\Phi_g}{\Phi_e}}\frac{\phi_g(z_T)\phi_e(z_T)}{\ovI{\phi_g}{\phi_e}}
\end{equation}
The first term denotes the relative change in the lattice constant while the second and third terms contain the dependence on the in-plane QD size and QD height, respectively. We find the functional dependence, $f$, between the third term and the quantization energy, $E-E_0$, using a finite-potential-well model, where $E$ is the emission energy and $E_0$ the bulk band gap of the QD material. We therefore obtain
\begin{equation}
\frac{\Lambda}{\mu} = S \times f(E-E_0),
\end{equation}
where $S=-(\Delta a_l/a_l) \me{\Phi_g}{x^2}{\Phi_e}/\ovI{\Phi_g}{\Phi_e}$. The trend of the experimental data from Fig.~\ref{fig:InterfaceData}(b), i.e., that $\Lambda/\mu$ increases with energy, agrees very well with our model (see the theory curve in the same figure), if the in-plane QD size is constant across the emission spectrum and only the height varies. This behavior has been reported in the literature in studies of QD size and shape using  similar growth conditions as used here~\cite{moison94}. We note that if the QDs had a constant aspect ratio, the mesoscopic strength would be predicted to decrease with energy because the effect depends stronger on the in-plane QD size, cf. \eqref{eq:Lambda_Mu_Gauss}. We note, that some degree of correlation between height and width has been observed~\cite{ebiko98}, but our study suggests that such correlations are small in our sample. We also stress that our study deals with the in-plane size of the QD wavefunctions, which generally can be different than the QD size measured by surface-profile techniques. We use $S$ and $E_0$ as fitting parameters and obtain a bulk band gap $E_0 = \SI{1.13}{\electronvolt}$ of the QD, which yields quantization energies ranging from \SI{40}{\milli\electronvolt} up to \SI{140}{\milli\electronvolt} across the emission spectrum. The resulting curve in Fig.~\ref{fig:InterfaceData}(b) agrees well with the experimental results.

\end{document}